\documentclass[12pt]{iopart}
\usepackage{graphicx} 
\usepackage{amssymb}
\usepackage{color}
\usepackage{changes}
\usepackage{framed}

\begin{document}

\title[Advanced  boundary electrode modeling for tES and parallel tES/EEG]{Advanced boundary electrode modeling for tES and parallel tES/EEG}

\author{Britte Agsten\dag,  Sven Wagner\dag, Sampsa Pursiainen\ddag, Carsten H. Wolters\dag}

\address{\dag Institute for Biomagnetism and Biosignalanalysis, Westf\"alische Wilhelms-Universit\"at M\"unster, Malmedyweg 15, 48149 M\"unster, Germany\\
\ddag Department of Mathematics, Tampere University of Technology, PO Box 553, FI-33101 Tampere, Finland}
\ead{sampsa.pursiainen@tut.fi}
\vspace{10pt}
\begin{indented}
\item[]March 2016
\end{indented}

\begin{abstract}
This paper explores advanced electrode modeling in the context of separate and parallel transcranial electrical stimulation (tES) and  electroencephalography (EEG) 
measurements. We focus on boundary condition based approaches that do not necessitate adding auxiliary elements, e.g.\ sponges,  to the computational domain. In particular, we investigate the complete electrode model (CEM) which incorporates a detailed description of the skin-electrode interface including its contact surface, impedance and normal current distribution. The CEM can be applied for both tES and EEG electrodes which is advantageous when a parallel system is used. In comparison to the CEM, we test two important reduced approaches: the gap model (GAP) and the point electrode model (PEM). We aim to find out the differences of these approaches for a realistic numerical setting based on the stimulation of the auditory cortex. The results obtained suggest, among other things, that GAP and GAP/PEM are sufficiently accurate for the practical application of tES and parallel tES/EEG, respectively. Differences between CEM and GAP were observed mainly in the skin compartment,
  where o
 nly CEM explains the  heating effects characteristic to tES. 
\end{abstract}
%
%
%
%
%

\section{Introduction}

Transcranial electrical stimulation (tES) is a non-invasive, inexpensive and easy-to-perform brain stimulation technique which modifies neural excitability \cite{nitsche2000}. Over the last decades, similar to other techniques like transcranial magnetic stimulation (TMS) \cite{CHW:Bar85,CHW:Gom2013}, tES became an important instrument in both neuroscientific research and medical therapy \cite{paulus2011,priori1998,miniussi,Hermann2013,Thut2011, butson2012}. It formed the basis for new therapies for diseases such as depression \cite{depressionz}, Parkinson's disease \cite{parkinsonz}, Alzheimer's disease \cite{alzheimerz}, stroke \cite{strokez} and  memory loss \cite{memoryz}. It was also shown that tES might help in the treatment of chronical pain \cite{pain2z}. Moreover brain stimulation techniques are important instruments in neuroscientific research as they give a new opportunity to reveal causal links between evoked brain activity and cognitive processes  \cite{Hermann2013,miniussi,Thut2011}.

In the conventional version of tES, two large electrode pads (sponges) are attached to the skin and  a low direct or alternating current (0.5--2 mA) is applied to the head \cite{paulus2011,priori1998}. This current penetrates partly to the brain \cite{CHW:Hol2006,CHW:Wag2014} where it can increase or decrease the cerebral excitability \cite{paulus2011,nitsche2000,priori1998}. 
In order to investigate the reaction of the brain to systematic changes in brain oscillations, researchers have recently started to use tES  in parallel with electroencephalography (EEG) measurements  (see e.g. \cite{helfrich,zaehle}). Among these approaches is  the high definition tES (HD-tES) \cite{CHW:Dat2013,CHW:Laa2013,CHW:Sch2015} in which, instead of two large pads, a larger number of smaller sized gel electrodes are used to target specific cortical structures.  Especially for HD-tES, but also for two sponge pad scenarios, computer optimization approaches and realistic head volume conductor modeling were proposed to achieve better focality and intensity in the target brain areas \cite{CHW:Dmo2011,strokez,CHW:Sch2015}. 

One of the first questions researchers are confronted to in tES modeling and its combination with EEG is the choice of the electrode model. 
This paper focuses on boundary condition based approaches.   
In particular,  we investigate the complete electrode model (CEM) \cite{purs1,ollikainen2000,CHW:Som92,cheng,phdvauk} which covers a comprehensive set of electrode boundary conditions and can be applied for both tES and EEG. The CEM incorporates a detailed description of the skin-electrode interface including its contact surface, impedance and normal current distribution also known as the shunting effect, i.e., current circulation on the contact surface,  which alters the underlying electric  potential \cite{purs1,ollikainen2000}.  
The CEM can be advantageous in contemporary tES simulations, for example, to replace saline soaked sponge electrode models \cite{tdcscem, datta,CHW:Wag2014,CHW:Sch2015}, and  especially in HD-tES in which multiple stimulation electrodes are used and where it delivers a more flexible approach since it does not necessitate modelling the sponges in the computational domain. In comparison to the CEM, we test two important reduced approaches, the gap model (GAP) and the classical point electrode model (PEM) which are applicable for tES and EEG, respectively. While both of these models ignore the shunting currents described above, they differ in the incorporation of size and form of the electrodes, which is ignored in the well-known PEM, but taken into account in the GAP. Since it has been investigated that size and form of stimulation electrodes have an impact on the focality of tES \cite{altse}, we do not use the PEM as a model for stimulation electrodes.    
  
In this study, we aim to find out the accuracy and differences of the present electrode approaches for realistic scenarios of tES and parallel tES/EEG. As an example case we use a realistic numerical setting which approximates the stimulation of the auditory cortex.  The results obtained suggest, among other things,  that GAP and GAP/PEM are sufficiently accurate for tES and parallel tES/EEG simulation studies, respectively. The significant differences between CEM and GAP were observed mainly in the skin compartment, where only CEM explains the heating effects characteristic to tES.  We also suggest that the conventional sponge models can be duplicated by the CEM with respect to the essential features of tES.  

This paper is organized as follows: The theory section \ref{sectheory} includes a brief review of the electrode models, and the methods section \ref{secmethods} gives a detailed description of the numerical experiments. Section \ref{secresults} presents the results which are then discussed in Section \ref{secdiscussion} and concluded in Section \ref{secconclusion}. 

\section{Theory}
\label{sectheory}

Let $\Omega$ be the head domain and $e_\ell$,  $\ell = 1, 2, \ldots, L$ a set of $L$ electrodes on its exterior boundary $\partial \Omega$ 
with surface contact area $\left|e_\ell \right|$ and potential $U_\ell$. The current applied to the $\ell$-th electrode is denoted by $I_\ell$. For an active (tES) and passive (EEG) electrode, it holds that $|I_\ell| \geq 0$ and $I_\ell = 0$, respectively. Following from the Kirchhoff's law, we assume that the total sum of the currents is zero, i.e.,  $\sum_{i=1}^L I_\ell =0$. In other words, we do not take into account small current losses that might exist. Furthermore, the divergence of the total  current density $\vec{J}$ in $\Omega$ is zero, that is, 
\begin{equation}0=\nabla \cdot \vec{J}=\nabla \cdot \left(\vec{J}^p-\sigma \nabla u\right) \quad  \hbox{or} \quad \nabla \cdot (\sigma \nabla u)  = \nabla \cdot \vec{J}^p \quad \hbox{in} \quad \Omega \label{eq1}\end{equation} 
with $u$ denoting the scalar electric potential field, $\sigma$ the conductivity distribution of the head and $\vec{J}^p$ the primary current density (neural activity) in the brain. Equation (\ref{eq1}) follows from the Maxwell's equations via the quasi-static approximation \cite{CHW:Plo67}, and it predicts the potential field  for both tES and EEG. The right-hand side of (\ref{eq1}) is relevant only with respect to the EEG measurements in which $\vec{J}^p$ is to be detected. Namely, the stimulation potential field can be obtained by setting the right-hand side to zero, i.e.,   
\begin{equation} 
\label{zero_eq}
\nabla \cdot \left( \sigma \nabla u \right) = 0.
\end{equation} 

\subsection{Complete electrode model}

In order to solve (\ref{eq1}), one can apply the following  CEM boundary conditions \cite{cheng}:
\begin{eqnarray}
\label{cemboundary} 0  =  \sigma \frac{\partial u}{\partial n}(\vec{x})  , & \quad   \hbox{in   }  \, \, \, \partial \Omega \backslash \cup_{\ell=1}^L e_\ell, \\
\label{cemboundary2} I_\ell  = \int_{e_\ell}{\sigma \frac{\partial u}{\partial n}(\vec{x}) dS}, & \quad \hbox{for }   \, \ell = 1, 2, \ldots,  L,\\
\label{cemboundary3}  {U_\ell}  =  u(x)  +\tilde{Z}_\ell \sigma \frac{\partial u}{\partial n}(\vec{x}) , & \quad  \hbox{for }   \, \ell = 1, 2, \ldots,  L, \, \, \,  \vec{x} \in e_\ell. 
\end{eqnarray}
The first one of these is the assumption that  no currents pass the part of the scalp that is not covered by electrodes.   
The second one states that the total current flux through the $\ell$-th electrode  equals to the applied current $I_\ell$.  According to the third one,   the $\ell$-th electrode voltage $U_\ell$ is the sum of the skin potential and the skin-electrode potential jump $\tilde{Z}_\ell \sigma \frac{\partial u}{\partial n}(\vec{x})$ in which  $\tilde{Z}_\ell$  (Ohm $\hbox{m}^2$)  is  a pointwise effective contact impedance (ECI). For simplicity, we  assume that ECI is of the form $\tilde{Z}_\ell = Z_\ell  |e_\ell|$ with $Z_\ell$ (Ohm) denoting the average contact impedance (ACI) of the electrode. Consequently, the integral form of (\ref{cemboundary3}) can be written as \begin{equation}
\label{u_ell}
 {U_\ell}  = \frac{1}{| e_\ell |} \int_{e_\ell} u \, dS  + {Z}_\ell { I_\ell }, 
\end{equation}
i.e., $U_\ell$ is the sum of the mean skin potential and the potential jump ${Z}_\ell { I_\ell }$. 
Additionally, the zero potential level is defined as $\sum_{I=1}^{L}U_\ell = 0$. 
Thus, the potential field $u \in S$ can be approximated by solving the following weak form (\ref{app1})
{\setlength\arraycolsep{2pt}\begin{eqnarray}
\textnormal{CEM: }\qquad\int_\Omega \sigma \nabla u \cdot \nabla v \, d V   =    \! \! &-& \int_\Omega (\nabla \cdot \vec{J}^p) v \, dV\nonumber \\
 &+& \sum_{\ell = 1}^L   \frac{I_\ell}{| e_\ell |}  \int_{e_\ell} v \, dS \nonumber \\  \! \!  &+&   \sum_{\ell = 1}^L \frac{1}{{Z}_\ell | e_\ell |^2} { \int_{e_\ell} u \, dS \! \int_{e_\ell}  v \, dS}  \nonumber \\  \!\! & - & \sum_{\ell = 1}^L\frac{1}{{Z}_\ell | e_\ell |}  \int_{e_\ell} u v \, d S \, \hbox{ for all } \,  v \! \in \!  \mathcal{S} \! \subset \! H^{1}(\Omega) \label{weak_form}
\end{eqnarray}}
in which $\mathcal{S}$ is a suitably chosen subspace of  $H^{1}(\Omega)$, i.e., the Sobolev space of square integrable functions with square integrable first-order partial derivatives. In this study, we use the finite element method to discretize $\Omega$. Consequently, $\mathcal{S}$ is assumed to be spanned by continuous finite element basis functions. In the weak form (\ref{weak_form}) of the CEM, the left-hand side defines a diffusion operator, the first two terms on the right-hand side correspond to neural and stimulation sources, respectively, and the third and fourth term describe the shunting effects: Due to their lower resistance currents tend to flow through the electrodes rather than the skin if the ACI is low enough (Figure \ref{fig1}). The higher the ACI, the weaker are the shunting effects. \\
Since currents distribute more freely underneath the electrode there is an equalizing effect on the potential underneath the electrode, which leads to a constant potential in the most extreme case (Figure \ref{fig2}).

\subsection{Reduced models}

In  GAP, the second CEM boundary condition (\ref{cemboundary2}) is replaced with the following pointwise formula 
 \begin{equation}
 \label{gap_boundary} \sigma \frac{\partial u}{\partial n}(\vec{x}) = \frac{I_\ell}{\left|e_\ell\right|} \quad \, \hbox{ for all }  \, \quad \vec{x} \in e_\ell, \quad  \ell = 1, 2, \ldots,  L. \end{equation} 
That is, the normal current density through the skin is assumed to be constant on each electrode (hence shunting effects are not taken into account). Furthermore, the third condition  (\ref{cemboundary3})  is  assumed to hold only in its integral form (\ref{u_ell}).  The resulting weak form is given by (\ref{app1})
\begin{eqnarray}
\label{weak_form_gap}
 \textnormal{GAP: }\qquad\int_\Omega \sigma \nabla u \cdot \nabla v \, d V  =     &-&\int_\Omega \! (\nabla \cdot \vec{J}^p) v \, dV\nonumber\\
 &+& \sum_{\ell = 1}^L   \frac{I_\ell}{| e_\ell |} \!  \int_{e_\ell} \! \! v \, dS \, \hbox{ for all } \, v \! \in \! \mathcal{S}. 
\end{eqnarray} 

The PEM boundary conditions follow from both the CEM and the GAP, if the support of each electrode tends to one point, i.e., $ e_\ell \to \vec{p}_\ell$ and $|e_\ell| \to 0$ for $\ell = 1, 2, \ldots, L$.  By taking the limit $\frac{1}{| e_\ell |} \int_{e_\ell} f \, dS   \to   f(\vec{p}_\ell)$, the $\ell$-th electrode voltage (\ref{u_ell}) is of the form  $U_\ell = u(\vec{p}_\ell) + I_\ell Z_\ell$ and the weak forms (\ref{weak_form}) and (\ref{weak_form_gap}) are now given by 
\begin{eqnarray}
\label{weak_form_pem}
 \textnormal{PEM: }\qquad\int_\Omega \sigma \nabla u \cdot \nabla v \, d V = &-&\int_\Omega (\nabla \cdot {J}^p) v \, dV \! \nonumber\\
&+& \! \sum_{\ell = 1}^L   {I_\ell}v(\vec{p}_\ell) \, \hbox{ for all } \, v \! \in \! \mathcal{S}.  
\end{eqnarray}

\begin{figure}[t]
	\centering
	\begin{tabular}{cc}
	\includegraphics[width=0.3\textwidth]{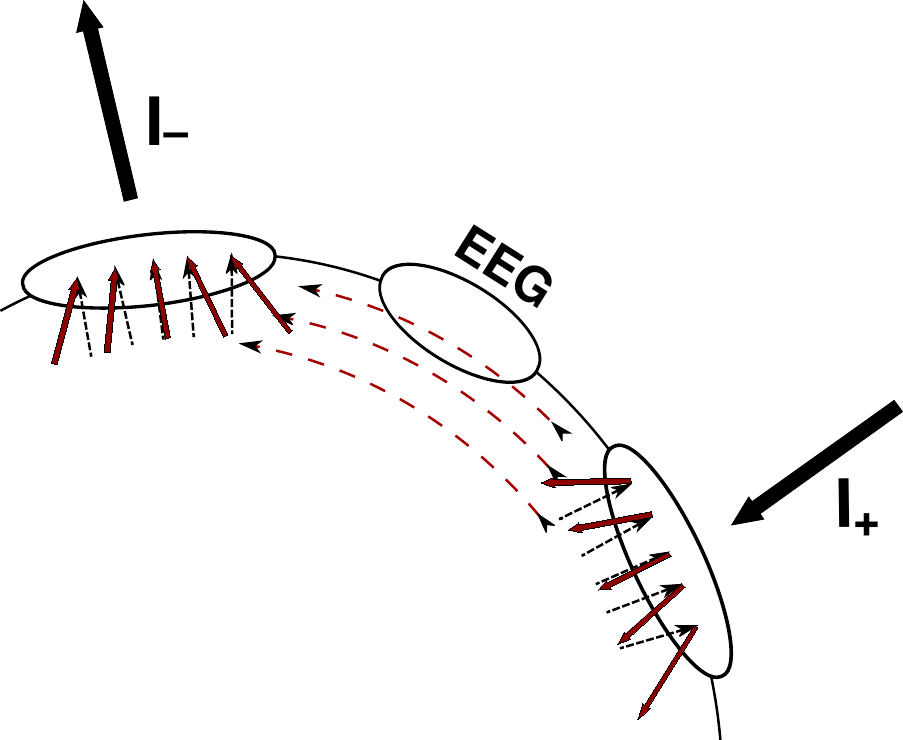}&
	\includegraphics[width=0.3\textwidth]{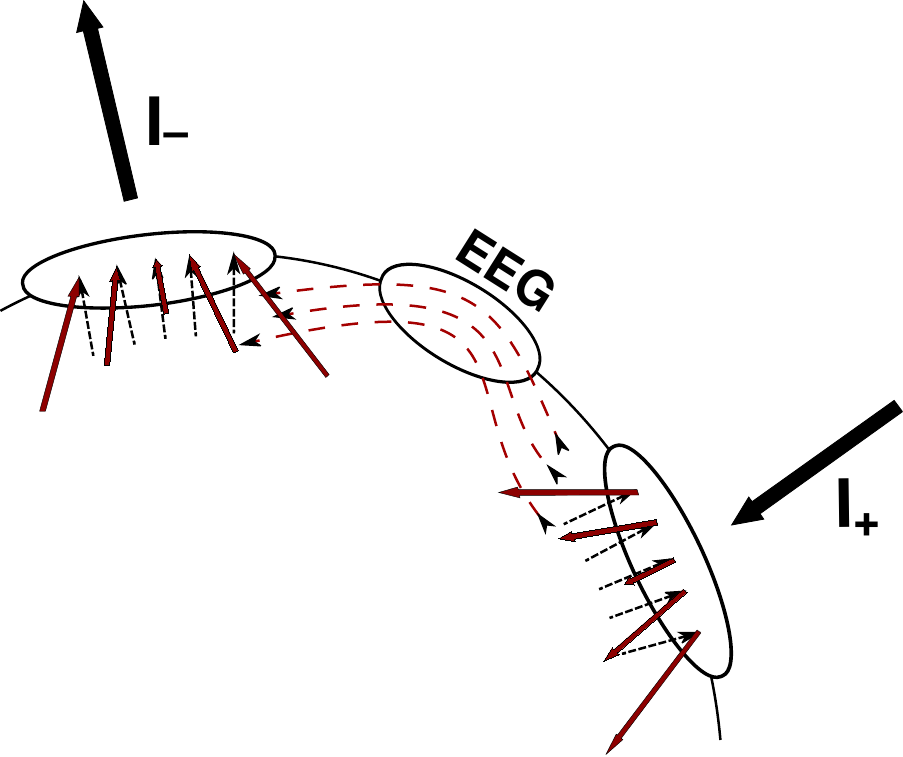}\\
	GAP/PEM& CEM 
	\end{tabular}
		\caption{Illustration of expected currents with and without shunting currents.}
		\label{fig1}	
\end{figure}

\begin{figure}[b]
	\centering
	\begin{tabular}{cc}
	 \includegraphics[width=0.4\textwidth]{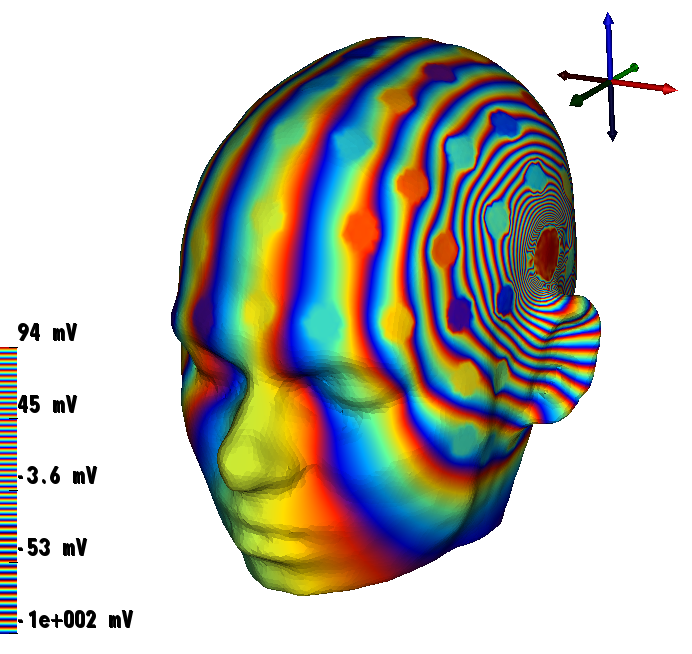}&
		\includegraphics[width=0.4\textwidth]{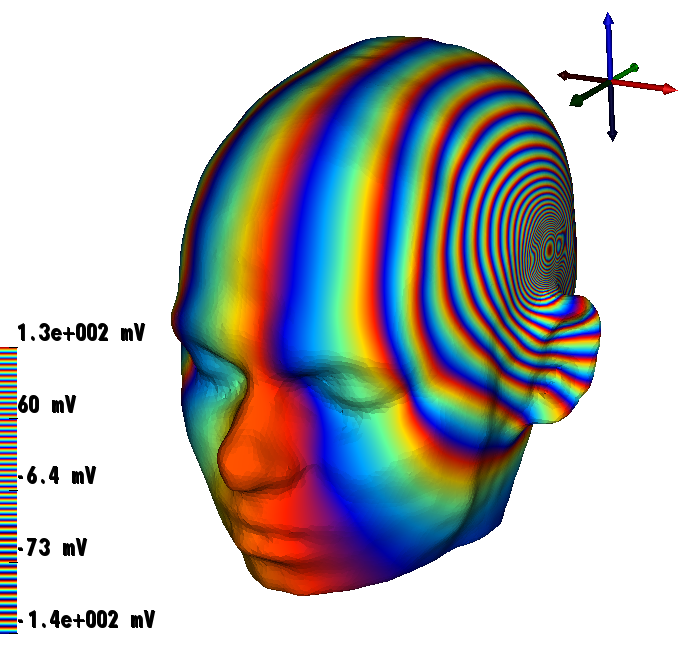}\\
		CEM (ACI 1 Ohm) & PEM/GAP
		\end{tabular}	
		\caption{Comparison of potential distributions in combined EEG and tES simulations, using different models. Anode and cathode are placed above the left and right ear. (\textit{Note that $1$ Ohm is a very low ACI which is used here to illustrate the effects of the CEM. For a more realistic ACI (e.g. $5$kOhm) the difference to
the GAP is hardly visible.})}
		\label{fig2}
\end{figure}

For the absence of the impedance-dependent terms of (\ref{weak_form}) in (\ref{weak_form_pem}) and (\ref{weak_form_gap}), i.e.\ for the omission of the shunting effects (Figure \ref{fig1}),  PEM and GAP can be interpreted as reduced electrode models compared to CEM. For $I_l=0$ the GAP model can be seen as the limit case of the CEM, where $Z_l$ is raised to infinity. Based on (\ref{weak_form_pem}) and (\ref{weak_form_gap}) it is obvious that for EEG  where the stimulation currents do not exist, GAP and PEM yield the same approximation for  $u$ up to a constant (zero-potential).  Furthermore, for small enough electrodes, each of the models CEM, PEM and GAP result in a similar approximation, that is, $CEM \approx PEM \approx GAP$.
 
\section{Methods}
\label{secmethods}
\subsection{Head model}
The head model of this study is based on T1- and T2- weighted magnetic resonance images (MRI) of a healthy 25-year old male subject that were registered accordingly and segmented into the most important head tissues \cite{purs1}. As a result of this procedure, non-intersecting surfaces of skin, skull compacta and spongiosa, brain grey matter and eyes were constructed. A constrained Delaunay tetrahedralization (CDT) was performed resulting in a tetrahedral mesh with 628,032 nodes and 3,912,563 tetrahedral elements. 
The conductivity values (in $Sm^{-1}$) for the different compartments were chosen to be 0.43 for skin \cite{CHW:Dan2011}, 0.0064 for skull compacta and 0.02865 for skull spongiosa \cite{CHW:Akh2002,CHW:Dan2011}, 1.79 for the CSF  \cite{CHW:Bau97}, 0.33 for
the brain \cite{CHW:Dan2011} and 0.505 for the eyes \cite{CHW:Ram2006}.

\subsection{Electrodes}
\begin{figure}[t]
\centering
	\includegraphics[width=0.4\textwidth]{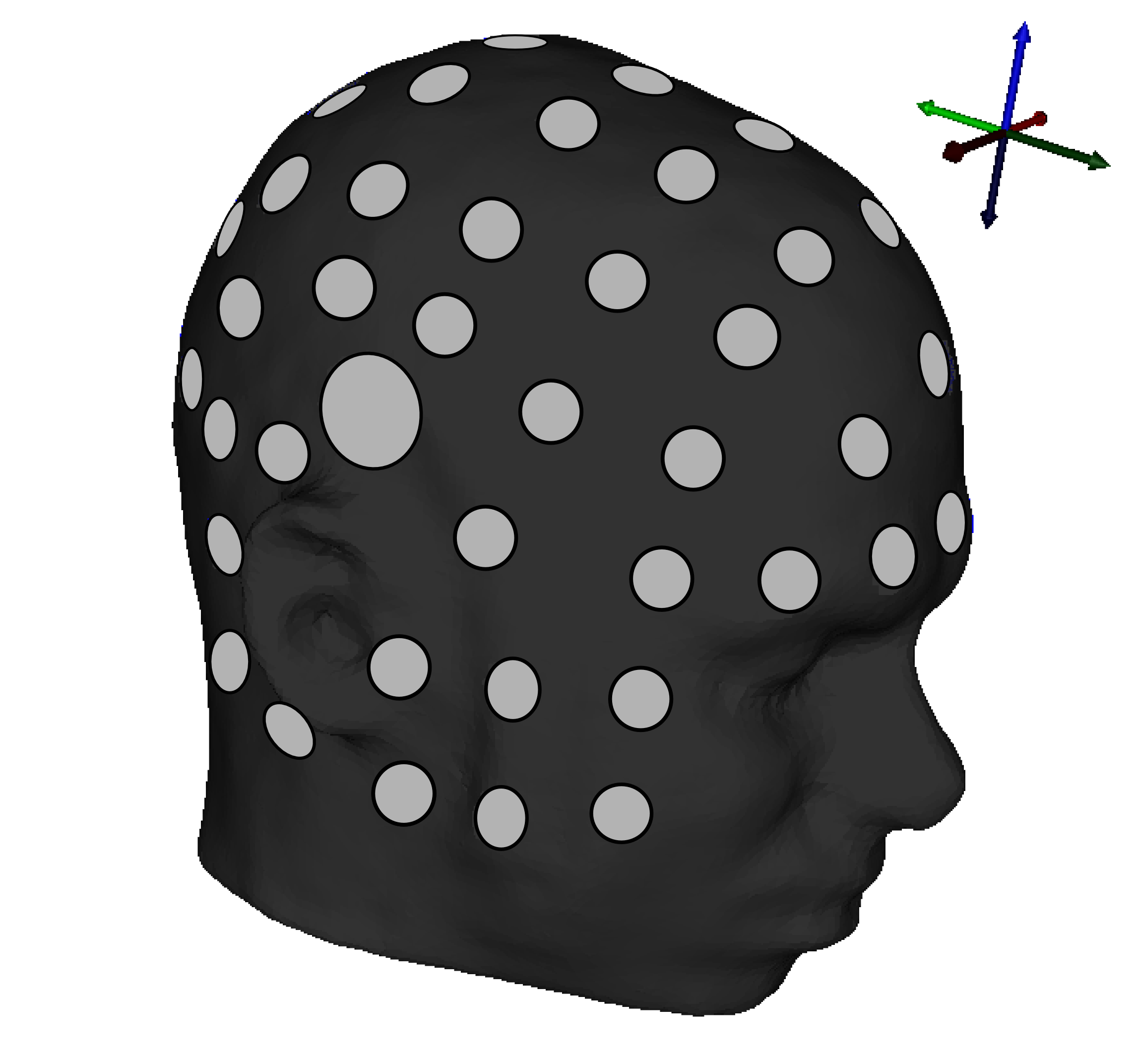} 
	\caption {Electrode configuration}
	\label{confi}	
	\end{figure}
We utilized a set of 79 electrodes which was positioned on the head surface according to the 10/10 system 
using the CURRY\footnote{http://compumedicsneuroscan.com/} software. The electrodes were formed as a combination of surface triangles (Figure \ref{confi}). The anode and cathode of tES were associated with two $24\,mm$ diameter electrodes placed over the ears similar to the stimulation of the auditory cortex. The EEG electrodes were assumed to have the average diameter  of $12 \, mm$. In all tests, we used a stimulation current of $1\,\mathrm{mA}$ injected and extracted above the left and right ear, respectively, i.e. through the tES anode and cathode. The stimulation potentials and currents within the head were computed using the tES/EEG electrode model combinations described in Table \ref{table:config}. Of these,  CEM/CEM and GAP/PEM were of the primary interest, CEM/PEM was tested in order to distinguish between the shunting of tES and EEG electrodes, and {$\hbox{CEM}_{1}$/$\hbox{CEM}_{1}$} was investigated in order to find out the effects of a very  low ACI. 
\begin{table}[t]
\caption{The electrode model combinations utilized for tES and EEG tested in the numerical experiments.}
\label{table:config}
  \begin{indented}
\item[]
\begin{tabular}{@{}lll}
\br
Name & tES model & EEG model \\
\mr
CEM/CEM  & CEM, ACI 5 kOhm & CEM, ACI 5 kOhm \\
{$\hbox{CEM}_{1}$/$\hbox{CEM}_{1}$} &  CEM, very low ACI 1 Ohm & {CEM}, very low ACI 1 Ohm \\
{GAP/PEM} & GAP & PEM  \\
{CEM/PEM} & CEM, ACI  5 kOhm & PEM    \\
\br
\end{tabular}
 \end{indented}
\end{table}

\subsection{Computation}
The head domain $\Omega$ was discretized using the finite element method.  For the linearity of (\ref{weak_form})  with respect to ${\bf I} = (I_1, I_2, \ldots, I_L)$,  the resulting linear system is of the form ${\bf x} = {\bf R}{\bf I}$ where ${\bf R}$ is a resistivity matrix and ${\bf x}$ is a discretization of $u$ (\ref{app2}). The final form of ${\bf R}$ depends on the finite element mesh, basis functions, as well as on the chosen electrode model.  In this study, we employed a tetrahedral mesh together with a set of first-order (nodal) Lagrange basis functions. A detailed description and derivation of ${\bf R}$ can be found in \cite{phdvauk,magsten}.
 After the calculation of $u$, the volume current density $-\sigma \nabla u$ in the brain was evaluated. In order to illustrate the differences between the investigated models, the following current angle and magnitude differences were evaluated for each tetrahedron:
\begin{equation} \hbox{Angle}(j_1,j_2)= \arccos\,\left(\frac{\left\langle j_1,j_2\right\rangle}{\left\|j_1\right\|\left\|j_2\right\|}\right), \quad \!\!\!  \hbox{Magnitude}(j_1,j_2)=\frac{\left\|j_1\right\|}{\left\|j_2\right\|}. \end{equation}

\section{Results}
\label{secresults}
\subsection{Current densities in head and brain} 	

\begin{table}[t]
\caption{Current extrema in brain $B$ and head $\Omega$ in $\frac{mA}{m^2}$.}
\label{table:currents}
  \begin{indented}
\item[]
\begin{tabular}{@{}llllll}
\br
 & \centre{2}{$B$}  &  \centre{2}{$\Omega$} \\
& \crule{2}  &  \crule{2} \\
& Min & Max &  Min & Max \\
\mr
 CEM/CEM & 3.052 & 89.52 &  7.5e-3& 4.217e4  \\
{$\hbox{CEM}_{1}$/$\hbox{CEM}_{1}$} & 2.870 & 83.26 &  5.6e-3& 5.812e4  \\
 {GAP/PEM} & 3.053 & 89.55 &  7.5e-3& 4.216e4 \\
 {CEM/PEM} &3.053 & 89.55 &  7.5e-3 & 4.217e4 \\
\br
\end{tabular}
 \end{indented}
\end{table}

\begin{figure}[b]
\centering
	\begin{tabular}{c c}
	 \includegraphics[width=0.3\textwidth]{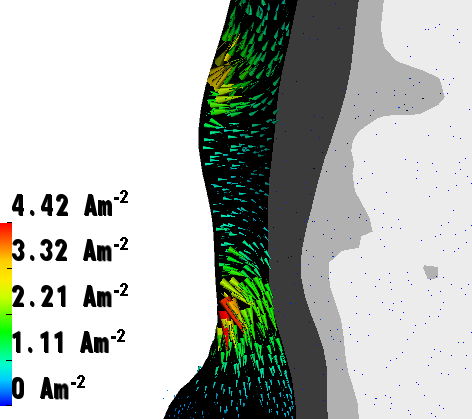}&
	\includegraphics[width=0.3\textwidth]{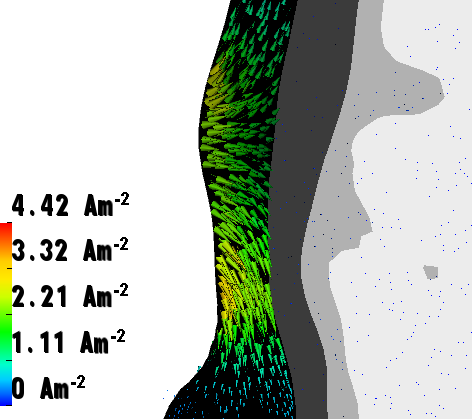}\\
	{CEM (ACI 1 Ohm)} & {GAP}
	\end{tabular}
	\caption[Visualization of shunting effect at the anode.]{Visualization of the surface current density at the anode. In CEM currents become stronger on the edges due to the shunting effect, i.e., current circulation on the contact surface, which is well visible only for low impedance value (ACI $1$ Ohm). In GAP, the normal current density is constant over the contact surface, meaning that the shunting effect is absent.}  
		\label{edgecurrents}
	\end{figure}
\begin{figure}[!]
\begin{center}
\begin{framed}
GAP/PEM \\
\begin{minipage}{5.3cm} 
\begin{center}
\includegraphics[width=5cm]{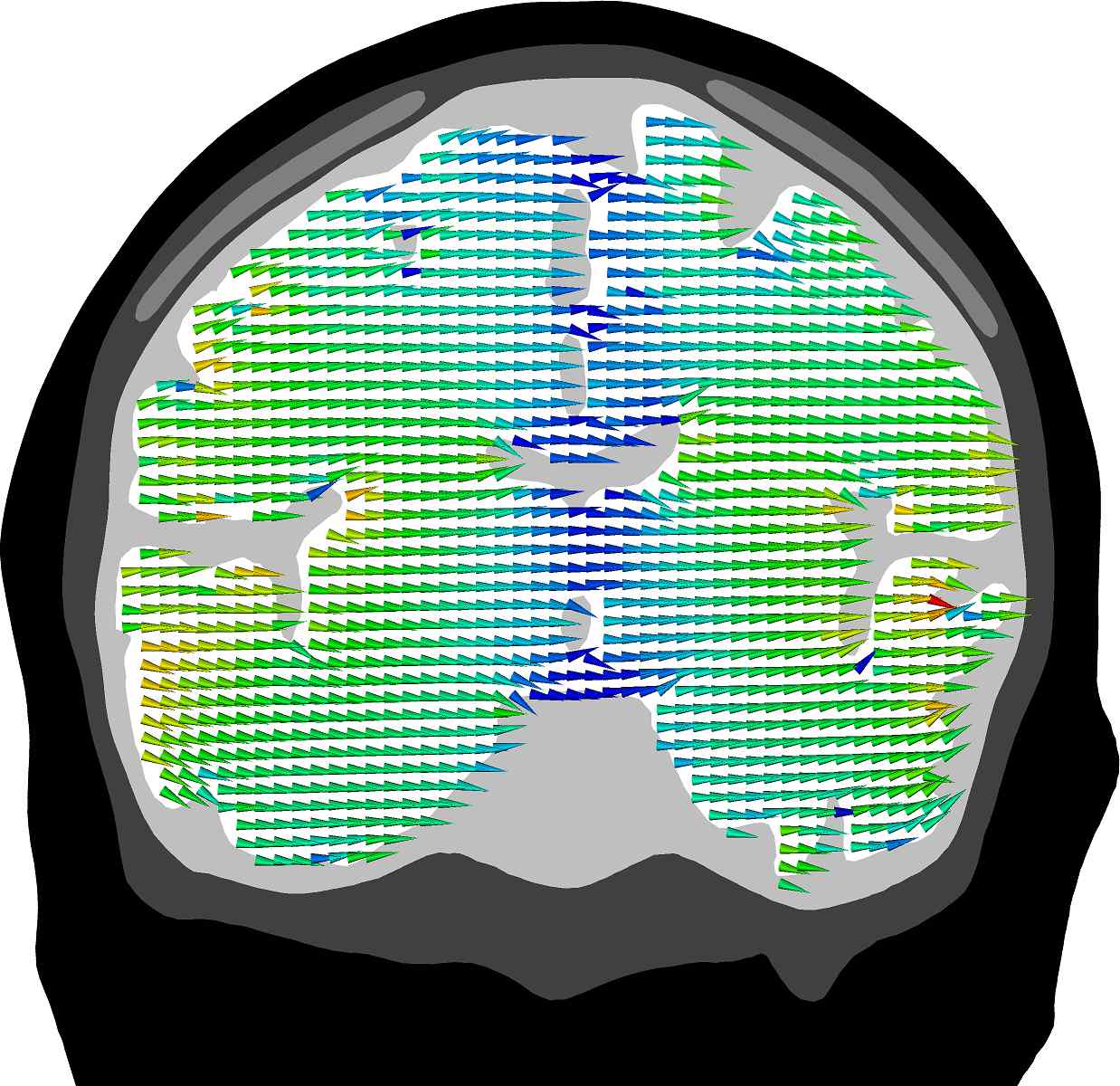} \\ 
Brain
\end{center}
\end{minipage}
\begin{minipage}{0.7cm} 
\begin{center}
\includegraphics[height=3.5cm]{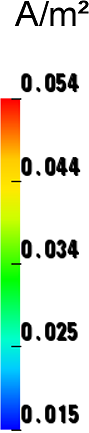} \\ 
\end{center}
\end{minipage}
\begin{minipage}{5.3cm} 
\begin{center}
	\includegraphics[width=5cm]{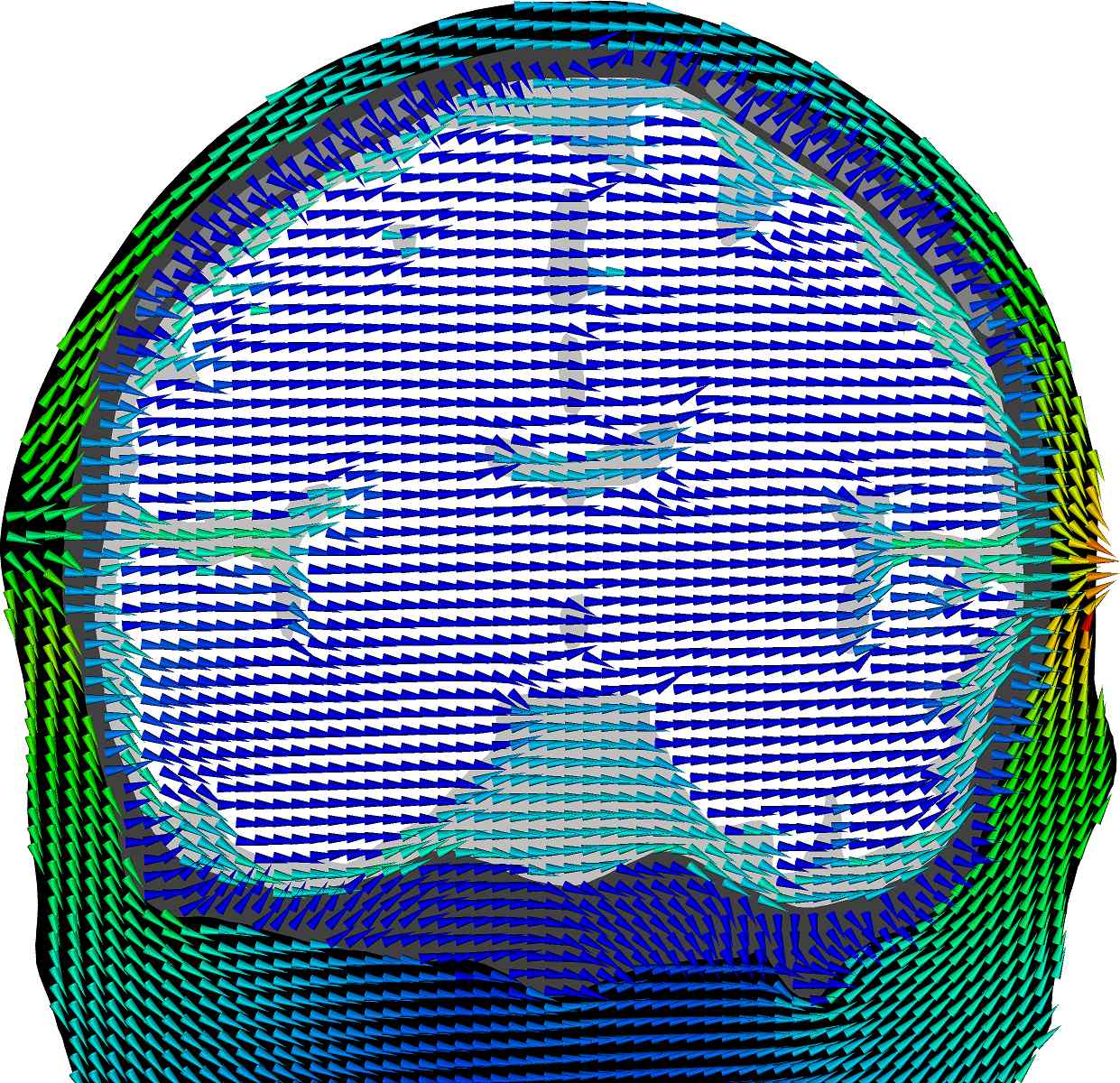}  \\
Head
\end{center}
\end{minipage} 
\begin{minipage}{0.7cm} 
\begin{center}
\includegraphics[height=3.5cm]{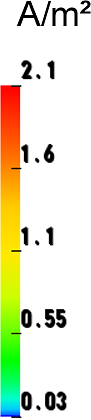} \\ 
\end{center}
\end{minipage}
\end{framed}
\begin{framed} 
$\hbox{CEM}_{1}$/$\hbox{CEM}_{1}$ \\
\begin{minipage}{5.3cm}  
\begin{center}
\includegraphics[width=5cm]{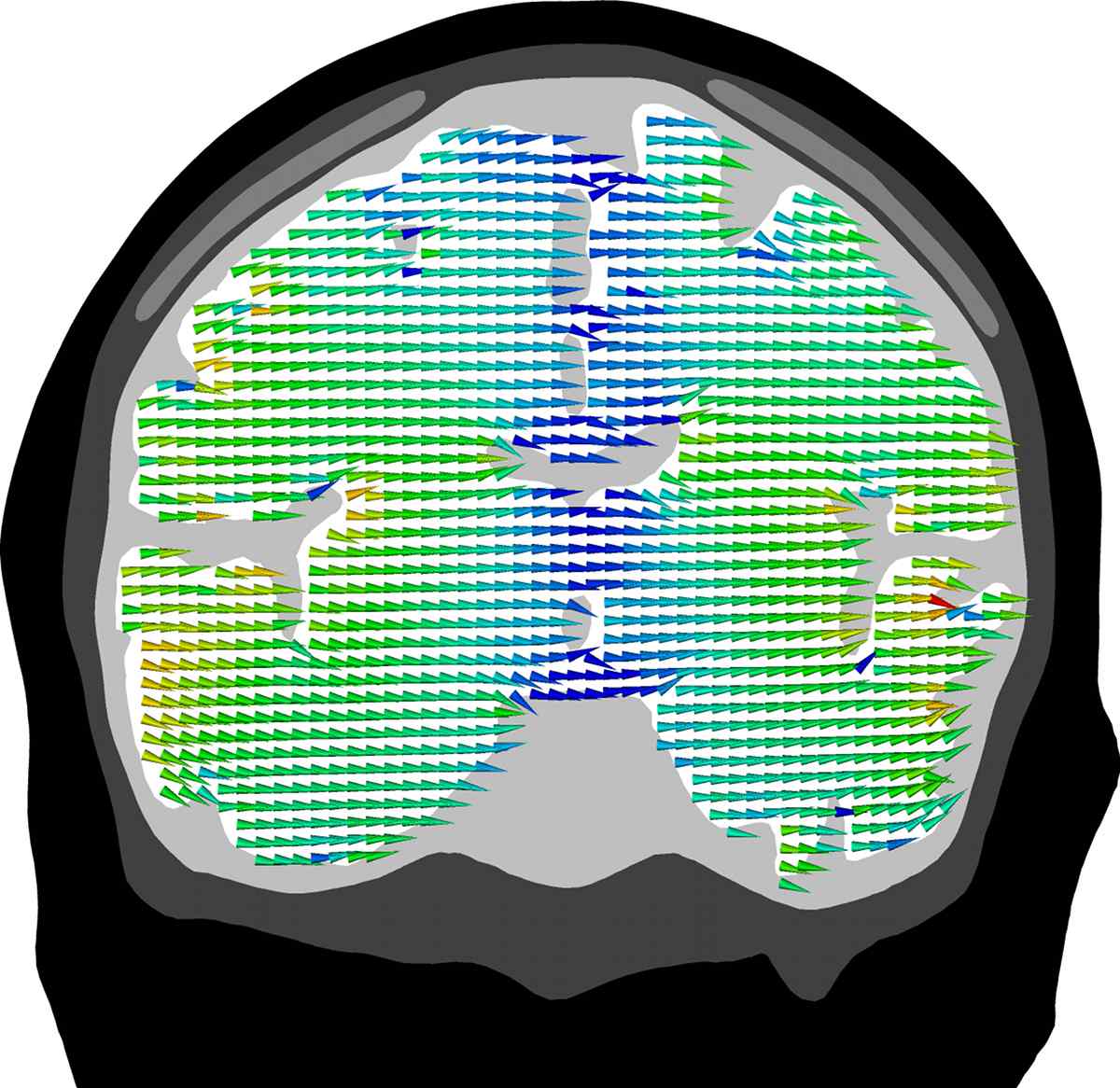} \\ Brain
\end{center}
\end{minipage}
\begin{minipage}{0.7cm} 
\begin{center}
\includegraphics[height=3.5cm]{brain_colorbar.png} \\ 
\end{center}
\end{minipage}
\begin{minipage}{5.3cm}  
\begin{center}
	\includegraphics[width=5cm]{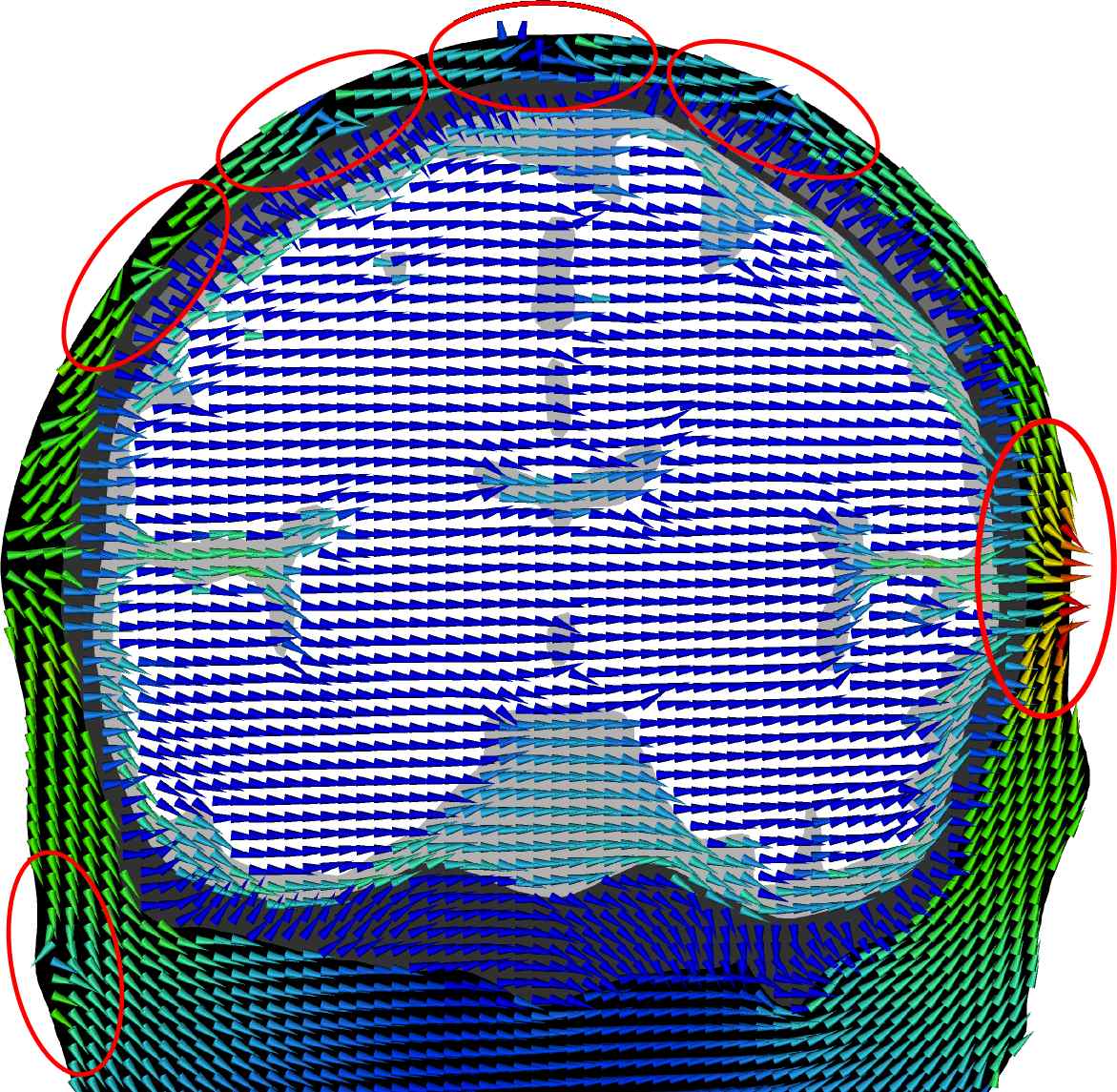} \\ Head
\end{center}
\end{minipage}
\begin{minipage}{0.7cm} 
\begin{center}
\includegraphics[height=3.5cm]{head_colorbar.png} \\ 
\end{center}
\end{minipage}
\end{framed}
\end{center}
	\caption{Currents in brain and head for GAP/PEM and $\hbox{CEM}_{1}$/$\hbox{CEM}_{1}$. Visible differences occur in the skin compartment under the electrodes (red ovals).  }
		\label{fig:Currents}
	\end{figure}	
Table \ref{table:currents} gives an idea of typical current strength in the brain and head compartments: One can observe that, with the exception of CEM$_{1}$/CEM$_{1}$, the differences are minor. Omitting CEM$_{1}$/CEM$_{1}$, the greatest differences in the maximal current density can be found between GAP/PEM and the other tested model combinations, indicating that those differences are related to the GAP (reduced) stimulation model. Compared to the GAP/PEM,  in models incorporating the CEM the maximum on the scalp is increased, while the maximum inside the brain is decreased. The reason for this phenomenon can be seen in Figure \ref{edgecurrents} which shows the current density on the anode for CEM and GAP. Although the effect is only shown for CEM$_{1}$/CEM$_{1}$ here, the same tendency can be expected for every value of the ACI (whereby a higher ACI leads to a weaker effect). As illustrated in the Figure, in CEM, the currents tend to concentrate on the edges of the electr
 odes, be
 cause the CEM allows for the current to distribute freely over the whole electrode. Since currents tend to flow through the (low impedance) skin rather than through the (high impedance) skull they will take the shortest way and leave the electrode at the edges. These strong currents are reflected by the maximum values in Table \ref{table:currents}. In contrast, the GAP forces the current density going through the electrode to be constant in the normal direction. However, the effects shown in Figure \ref{edgecurrents} are not as strong for the realistic CEM, and therefore differences due to this phenomenon are small. 

Figure \ref{fig:Currents} shows the currents in brain and head for GAP/PEM and CEM$_{1}$/CEM$_{1}$. It is visible here again that the  injected current tends to flow through the skin instead of the brain, where only a minor part of the current  is led to. The current patterns show that effects can mainly be expected under the electrodes, whereas the current flow in the head is unaffected even for low impedance values.
\subsection{Angle and magnitude differences}\label{sec:anglemag}
\begin{table}[t]
\caption{Maximal and minimal angle  magnitude differences in brain and head.}
\label{table:magnitude}
\begin{indented}
\item[]
\begin{tabular}{@{}llllll}
\br
& & \centre{2}{$B$}  &   \centre{2}{$\Omega$} \\
& & \crule{2}  &  \crule{2} \\
& & Min & Max &  Min & Max \\
\mr 
Angle ($\mbox{}^\circ$) &{CEM/CEM vs.\ GAP/PEM} & $\sim 0$  & \- 0.040 &  $\sim 0$ & 1.04  \\
 &{CEM/PEM} vs.\  {GAP/PEM} & $\sim 0$ &  \- 0.041 &  no diff. & 0.21 \\
 &{CEM/CEM} vs.\ {CEM/PEM} &$\sim 0$  &  \- 0.024 &  no diff. & 1.04\\ \mbox{} \\
Mag.\ (\%) & {CEM/CEM} vs.\ {GAP/PEM} & -0.086 & \- 0.010 &  -1.71& 1.31  \\
&  {CEM/PEM} vs.\ {GAP/PEM} & -0.041 & \- 0.036 &  -0.46 & 0.35 \\
 & {CEM/CEM} vs.\ {CEM/PEM} &-0.059 &  -0.003 &  -1.72 & 1.30 \\ \br
\end{tabular}
\end{indented}
\end{table}
Angle and magnitude differences were evaluated in order to get exact numerical data of the modeling characteristics. Maximal and minimal angle and magnitude differences are presented in Table \ref{table:magnitude}. As expected, the  currents differ the most when the CEM/CEM is compared to GAP/PEM, but even then differences lie below $2^{\circ}$ and $2\,\%$, respectively. In case of the brain, which is the primary target of the tES  stimulation, the differences are below $0.05^{\circ}$ and $0.1 \, \%$.
Overall, differences in the brain are sufficiently small to be ignored in all cases.\\
Changing the model at the stimulation electrodes (CEM/PEM vs. GAP/PEM) has a bigger influence on the brain than a model change at the passive electrodes (CEM/CEM vs. CEM/PEM), while it is the other way around on the scalp.

\begin{figure}[!]
\begin{center}
\begin{framed}
Angle differences ($\mbox{}^\circ$) \\
\begin{minipage}{5.3cm} 
\begin{center}
\includegraphics[width=5cm]{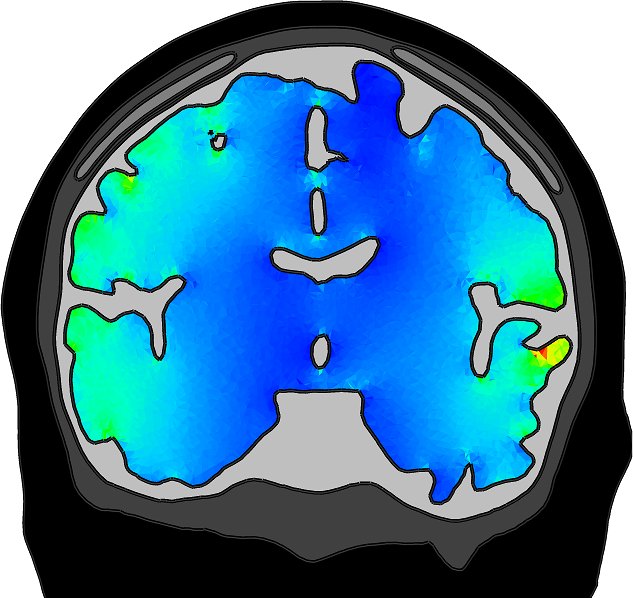} \\ 
Brain
\end{center}
\end{minipage}
\begin{minipage}{0.7cm} 
\begin{center}
\includegraphics[height=3cm]{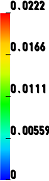} \\ 
\end{center}
\end{minipage}
\begin{minipage}{5.3cm} 
\begin{center}
	\includegraphics[width=5cm]{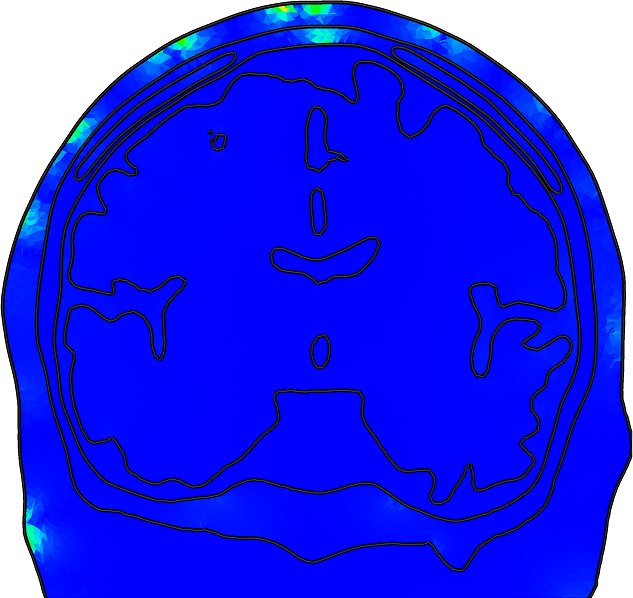}  \\
Head
\end{center}
\end{minipage} 
\begin{minipage}{0.7cm} 
\begin{center}
\includegraphics[height=3cm]{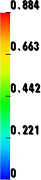} \\ 
\end{center}
\end{minipage}
\end{framed}
\begin{framed} 
Magnitude differences (\%)\\
\begin{minipage}{5.3cm} 
\begin{center}
 \includegraphics[width=5cm]{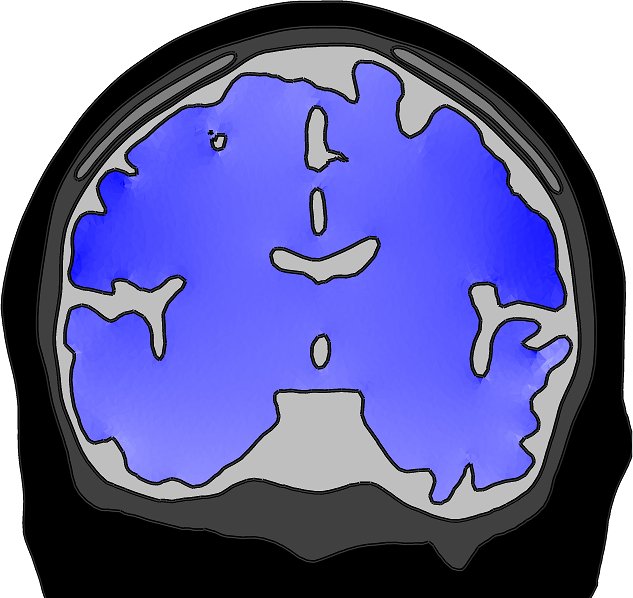} \\ Brain
\end{center}
\end{minipage}
\begin{minipage}{0.7cm} 
\begin{center}
\includegraphics[height=3cm]{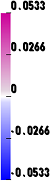} \\ 
\end{center}
\end{minipage}
\begin{minipage}{5.3cm} 
\begin{center}
	\includegraphics[width=5cm]{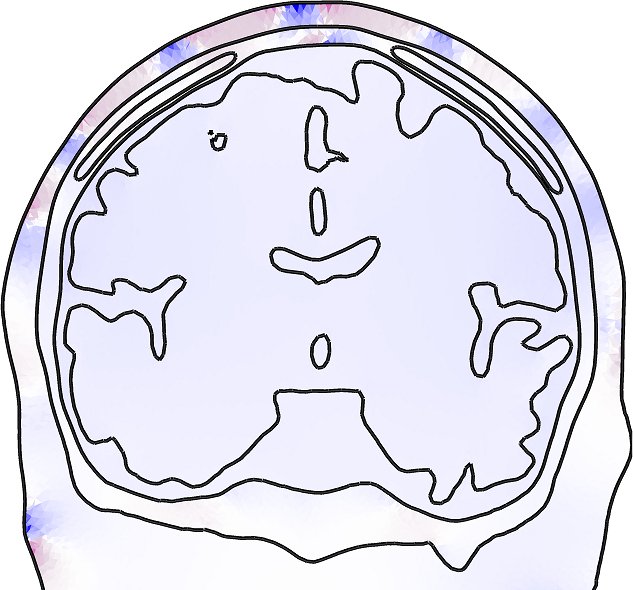} \\ Head
\end{center}
\end{minipage}
\begin{minipage}{0.7cm} 
\begin{center}
\includegraphics[height=3cm]{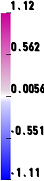} \\ 
\end{center}
\end{minipage}
\end{framed}
\end{center}
	\caption{Comparison of CEM/CEM and GAP/PEM in brain and head}
	\label{fig:Currentdiffs}
	\end{figure}

Figure \ref{fig:Currentdiffs} visualizes the distribution of angle and magnitude differences in the head and brain using the example of CEM/CEM vs.\ GAP/PEM. One can observe here as well, that differences occur mainly in the skin compartment, but hardly in the brain. Furthermore, differences in the skin compartment are mainly visible next to electrodes, which corresponds to the expectations, as differences will mainly reflect shunting effects. However, this also means that shunting effects will mainly have a local influence, and differences in the other parts of the head will be much lower.

Although differences in the brain are very low, the patterns in the brain show that differences occur mainly near to the stimulation electrodes. This supports the observation that the shunting effects at the stimulating electrodes have a bigger influence on the brain than those at the measurement electrodes. 

\section{Discussion}
\label{secdiscussion}
In this study, we compared different boundary electrode modeling approaches for forward simulation in tES and parallel tES/EEG. These models include the complete electrode model (CEM) and two important reduced approaches: the point electrode model (PEM) and the so-called gap model (GAP). Different combinations of these were tested for tES/EEG electrodes. The main emphasis was on the comparison of CEM/CEM and GAP/PEM. Additionally, CEM/PEM was tested in order to investigate the influence of EEG induced noise, and the effects of low ACI were explored via the combination CEM$_{1}$/CEM$_{1}$, where  CEM$_{1}$ was a CEM approach with a low ACI.  

It was observed that the CEM/CEM will not lead to significant orientation differences in the current density compared to GAP/PEM. This holds especially for the brain, where differences are shown to be below $0.05 ^\circ$ and magnitude differences below $0.1\%$, suggesting that there are no relevant differences in current distribution. The highest differences occur near to the electrodes themselves, but even here highest differences lie under $1.1^\circ$ and $2\%$, respectively. Hence, for stimulation focality it will not be relevant which of the presented models is used.  Moreover, based on the results,  orientation differences in the brain occur mainly due to stimulation electrodes. 

As stimulation approaches we investigated  the CEM and GAP. The PEM was not covered, since it is a well known fact that the size of stimulation electrodes has a strong impact on focality and current magnitudes \cite{altse}. One can therefore assume that the use of the PEM in tES does not make much sense, especially if large stimulation electrodes are used. 
This observation is supported by the applied studies which typically incorporate the size of the electrodes  \cite{tdcscem, datta}. 
Hence,  it is also obvious that the PEM can be used for EEG analysis, where it is currently the standard model. 

The shunting effects present in the CEM were observed to have the biggest impact in the skin compartment. It is well-known that, in tES, the skin-electrode current density  tends to be stronger on the edges of the stimulation electrodes and therefore can heat (burn) the skin. Our study confirms that the shunting effects have an impact on this tendency as the magnitude of edge currents is increased for the CEM. Hence, the CEM could also be an appropriate model to investigate the heating of the skin. The CEM may allow one, via pointwise control of the ECI, to design electrodes that yield more equal current distribution under the contact patch than the existing alternatives which can lead to heating. Such a  design could also serve the general goal to focus the stimulation currents to given areas within the brain, avoiding nuisance currents in other parts of the head, e.g., in the skin.

The boundary electrode models of this study are non-standard approaches in  tES although the PEM was sometimes used for stimulation in the past \cite{CHW:Suh2009}. Instead, it is very common to model electrodes as sponges with the  conductivity of saline \cite{CHW:Wag2014}. This is due to the fact that stimulation electrodes in tES are indeed often large saline soaked sponges \cite{tdcscem}.  Nevertheless, the sponge model has also been applied to disc electrodes in combination with saline gel \cite{datta, CHW:Dat2013}. In the sponge model,  the current is assumed to be applied through the sponges, and it is supposed that either the resulting potential \cite{tdcscem}  or the current density is constant over the whole electrode \cite{CHW:Wag2014, Sadlier2010}. In terms of the presented boundary electrode models, one can argue that the latter case, i.e., the assumption of constant current densities over the whole electrode, describes the application of the GAP, whereas the form
 er case 
 describes the CEM with a vanishing contact impedance, i.e., a model with maximal shunting currents. However, it is more likely that both approaches lead to results similar to a classical CEM, as the sponge can be seen as a representation of the shunting electrode. 

An important future work direction is to investigate the actual differences between the CEM and both versions of the sponge model. In such a study one will need to take into account that the actual shape of the sponge is not well-known in all cases, especially when saline gel is used  instead of a classical sponge \cite{CHW:Dat2013}. On the other hand, contact impedances on the electrodes can be measured exactly, which is why the CEM can be advantageous. In addition, the CEM does not necessitate adding sponges to the computational model and thus the mesh is more flexible and easier to construct. For these aspects, it is necessary to compare the accuracy of both approaches pertaining to real data. Another future direction is to model HD-tES stimulation patterns in parallel  EEG/tES using  the presented boundary condition based approach instead of the sponge model. In such a context, small high-definition electrodes can be used  for both tES and EEG and several stimulating patt
 erns can
  be investigated using a single electrode configuration \cite{datta}.
 Consequently, the resistivity matrix could  be studied as a whole in order to cover all possible linearly independent current patterns. Here, our results suggest that the gap model might be sufficient enough to replace both CEM and sponge model. For their intensity in the skin compartment, finding the significance of the shunting effects with respect to the density of the electrode configuration is also an interesting research topic.

\section{Conclusion}
 \label{secconclusion}

Advanced boundary condition based electrode modeling is an interesting alternative for tES and parallel tES/EEG studies. The present results suggest that the shunting effects present in the CEM are minor regarding practical brain stimulation applications. The reduced approaches GAP and PEM were found to be sufficient for modeling EEG and tES electrodes, respectively. The shunting effects present in the CEM were observed to have the biggest impact in the skin compartment.  Interesting future directions motivated by this study  include  the exploration of the CEM in comparison with the sponge models of tES, for variable current patterns and dense electrode configurations of HD-tES as well as for avoiding nuisance electrode currents.

\section*{Acknowledgement}

This work was partially supported by the Priority Program 1665 of the Deutsche Forschungsgemeinschaft (DFG) (WO1425/5-1), the EU project ChildBrain (Marie Curie Innovative Training
Networks, grant agreement no. 641652), Academy of Finland project () and the Academy of Finland Center of Excellence in Inverse Problems.

\appendix
\section{Integration by parts}
\label{app1}
Multiplying (\ref{eq1}) with a test function $v \in S$ and integrating by parts yields
{\setlength\arraycolsep{2pt} \begin{eqnarray} \label{partial_integration}
-\int_\Omega (\nabla \cdot {J}^p) v \, dV & = &   -\int_\Omega \nabla \cdot( \sigma \nabla u) v  \, d V  \nonumber \\ & =&  \int_\Omega \sigma \nabla u \cdot \nabla v \, d V - \int_{\partial \Omega}  (\sigma \frac{\partial u}{\partial n}) v \, d S \nonumber \\ & = &  \int_\Omega \sigma \nabla u \cdot \nabla v \, d V  - \sum_{\ell = 1}^L  \int_{e_\ell}    ( \sigma \frac{\partial u}{\partial n}) v \, d S.
\end{eqnarray}}
On the last step we have used (\ref{cemboundary}). Under the pointwise GAP boundary condition (\ref{gap_boundary}) this directly yields the weak form of GAP (\ref{weak_form_gap}). The CEM weak form (\ref{weak_form}) can be obtained as follows: 
{\setlength\arraycolsep{2pt} \begin{eqnarray}
 - \sum_{\ell = 1}^L  \int_{e_\ell}    ( \sigma \frac{\partial u}{\partial n}) v \, d S  & = &  - \sum_{\ell = 1}^L  \frac{1}{{Z}_\ell | e_\ell |} \int_{e_\ell}  (U_\ell - u) v \, d S  \nonumber \\
& =& -\sum_{\ell = 1}^L  \frac{U_\ell}{{Z}_\ell | e_\ell |} \int_{e_\ell}  v \, dS  +  \sum_{\ell = 1}^L  \frac{1}{{Z}_\ell | e_\ell |}  \int_{e_\ell} u v \, d S \nonumber \\ & = & - \sum_{\ell = 1}^L \frac{I_\ell}{| e_\ell |}    \int_{e_\ell} v \, dS  - \sum_{\ell = 1}^L   \frac{1}{{Z}_\ell | e_\ell |^2} { \int_{e_\ell} u \, dS\int_{e_\ell}  v \, dS}  \nonumber \\ & & +  \sum_{\ell = 1}^L  \frac{1}{{Z}_\ell | e_\ell |}  \int_{e_\ell} u v \, d S .
\end{eqnarray}}
Substituting the final form of the right-hand side into (\ref{partial_integration}) leads to  (\ref{weak_form}).

\section{Resistivity matrix}
\label{app2}

In this section, we derive the resistivity matrix ${\bf R}$ for CEM, PEM and GAP in the case of the  zero right-hand side potential equation (\ref{zero_eq}). 
Given scalar valued basis functions $\psi_1, \psi_2, \ldots, \psi_{n} \in \mathcal{S}$, the  potential $u$ can be approximated as the finite sum $u = \sum_{i = 1}^{N} x_i
\psi_i$. Denoting by ${\bf x} = (x_1, x_2,\ldots, x_{N})$ the coordinate vector of the discretized potential and by ${\bf Y} = (Y_1, Y_2,\ldots, Y_{L})$ the (ungrounded) electrode voltages,  the CEM weak form  (\ref{weak_form})  is given by 
 \begin{equation}
\label{u_system} \left( \begin{array}{cc} {\bf A} & -{\bf B} \\
-{\bf B}^T & {\bf C}
\end{array} \right) \left( \begin{array}{c} {\bf x}  \\
{{\bf Y}}
\end{array} \right) = \left( \begin{array}{c} {\bf 0}  \\
{\bf I}
\end{array} \right).
\end{equation}
Here, ${\bf A}$ is of the form
\begin{equation}
a_{i, j}   =  \int_{\Omega} \sigma \nabla \psi_i \cdot \nabla \psi_j \, d V +
\sum_{\ell = 1}^L \frac{1}{Z_\ell |e_\ell|} \int_{e_\ell} \psi_i \psi_j \, dS, 
\label{tupu}
\end{equation}
and the entries of ${\bf B}$ ($N$-by-$L$) and ${\bf C}$ ($L$-by-$L$)  are given by 
{\setlength\arraycolsep{2pt}
\begin{eqnarray} 
\label{fem_system} 
b_{i, \ell}  & = & \frac{1}{Z_\ell |{e}_\ell|} \int_{e_{\ell}} \psi_i \, dS  , \label{hupu} \\
c_{\ell, \ell}  & = & \frac{1}{Z_\ell}.
\end{eqnarray}} 
Consequently, the resistivity matrix satisfying ${\bf x} = {\bf R} {\bf I}$ can be expressed as
\begin{equation}
\label{resistivity} {\bf R}  =   {\bf A}^{-1} {\bf B} ( {\bf C}  - {\bf B}^{T} {\bf A }^{-1} {\bf B} )^{-1} .
\end{equation} . 
In the case of the GAP, we replace ${\bf A}$ by the simplification  ${\bf A'}$ with $a'_{i, j}   =  \int_{\Omega} \sigma \nabla \psi_i \cdot \nabla \psi_j \, d V$ and, for invertibility, 
additionally $a'_{i',i'} = 1$ and $a'_{i',j} = 0$ for $j \neq i'$ for a single basis function $\psi_i'$ attaining its maximum on the part of the boundary not covered by the electrodes. Then the GAP weak form (\ref{weak_form_gap}) is given by 
\begin{equation}
{\bf A'x}={\bf BC}^{-1}{\bf I}
\end{equation}
Thus, the resistivity matrix is of the form
\begin{equation}
\label{resistivitygap}{\bf R} = {\bf A }'^{-1} {\bf B} {\bf C}^{-1}.
\end{equation}
The electrode potentials ${ \bf Y}$ can still be obtained by ${\bf-B^Tx+CY}={\bf I}$ (cf. second row of \ref{u_system}).

The resistivity matrix for the PEM weak form (\ref{weak_form_pem}) can be followed from both (\ref{resistivity}) and (\ref{resistivitygap})  by taking the limit   $\left|e_\ell\right| \to \vec{p}_\ell$ which leads to
\begin{equation}
b_{i, \ell}    \to  \frac{1}{Z_\ell} \psi_i(\vec{p}_\ell) 
\end{equation}
and 
\begin{equation}
a_{i, j}    \to   \int_{\Omega} \sigma \nabla \psi_i \cdot \nabla \psi_j \, d V +
\sum_{\ell = 1}^L \frac{1}{Z_\ell}  \psi_i(\vec{p}_\ell) \psi_j(\vec{p}_\ell) 
\end{equation}
or 
\begin{equation}
a'_{i, j}    \to   \int_{\Omega} \sigma \nabla \psi_i \cdot \nabla \psi_j \, d V. 
\end{equation}

In each case, the potentials must be grounded in order to make them comparable.  However, for comparing the volume current distributions $-\sigma \nabla u$ that is not necessary.

\section*{References}
\bibliography{literatur2} 
\bibliographystyle{abbrv}

\end{document}